\documentclass[review]{elsarticle}

\usepackage[breaklinks]{hyperref}
\usepackage{makecell}
\usepackage{float}
\usepackage{pifont}
\usepackage{xcolor}
\usepackage{balance}
\usepackage{multirow}
\usepackage{comment}
\usepackage{adjustbox}
\usepackage{url}
\usepackage{breakurl}

\journal{Journal of \LaTeX\ Templates}

\newcommand{\cmark}{\ding{51}}%
\newcommand{\xmark}{\ding{55}}%
\newcommand{\sol}{Maildust}

\newcolumntype{C}[1]{>{\centering\arraybackslash}p{#1}}

\begin{document}

\begin{frontmatter}

\title{
A Longitudinal Study on Web-sites 
Password Management (in)Security: Evidence and Remedies}
\author{Simone Raponi, Roberto Di Pietro}

\address{Division of Information and Computing Technology\\ College of Science and Engineering, Hamad Bin Khalifa University \protect\\ Doha, Qatar \\ sraponi@mail.hbku.edu.qa, rdipietro@hbku.edu.qa}

\begin{abstract}
Single-factor password-based authentication is generally the norm to access on-line Web-sites.
While single-factor authentication is well known to be a weak form of authentication, a further 
concern arises
when considering the possibility for an attacker to recover the user passwords by leveraging the loopholes in the password recovery mechanisms. Indeed, the adoption by a Web-site of a poor password management system makes useless even the most robust password chosen by the registered users.
In this paper, building on the results of our previous work, we study the possible attacks to on-line password recovery systems analyzing the mechanisms implemented by some of the most popular Web-sites. In detail, we provide several contributions: (i) we revise and detail the attacker model; (ii) we provide an updated analysis with respect to a preliminary study we carried out in December 2017; (iii) we perform a brand new analysis of the current top 200 Alexa's Web-sites of 
five major EU countries; 
and, (iv) we propose \sol, a working open-source  module that could be adopted by any Web-site to provide registered users with a password recovery mechanism to prevent mail service provider-level attacks.
Overall, it is striking to notice how
the  analyzed Web-sites have made little (if any) effort to become compliant with the GDPR regulation, 
showing that the objective to have basic user protection mechanisms in place---despite the fines threatened by GDPR---is still far, mainly because of sub-standard security management practices. Finally, it is worth noting that while this study has been  focused on EU registered Web-sites, the proposed solution has, instead, general applicability.
\end{abstract}

\end{frontmatter}

\section{Introduction}
Although online Web-sites have always been attractive targets for attackers, in the last years the phenomenon has intensified, typically leveraging the poor password management implemented on the Web-site~\cite{website-security-statistics-report}. 
The victims of these oversights are the registered users, that
find themselves having to quickly change the credentials, as a generic leakage compromised the currently used. A user could be asked to be compliant with the well-known best practices to choose a strong password but, if Web-sites' password storage does not respect basic security guidelines, such effort would be fruitless. LinkedIn, the professional networking Web-site, and Yahoo, the Web-service provider, are two victims of major data breaches in the last years~\cite{linkedin-data-breach, yahoo-data-breach}. The former, after having found 6.5 million encrypted passwords of its users on a Russian Web-site in June 2012, in May 2016 suffered another major breach that led to 117 million credentials for sale on the Dark Web~\cite{linkedin-dark-web}. Although it was a Web-site that enjoyed an excellent reputation with millions of active registered users, LinkedIn was implementing poor cryptography techniques to protect the credentials of the users~\cite{kamp2012linkedin}.
Regarding Yahoo, the company suffered one of the biggest data leakage ever (i.e., 3 billion users passwords exposed). It all started with a malicious phishing e-mail sent by the attacker to an employee of the company. After an inattentive click of the victim, a backdoor was installed in the internal system, that granted the attacker access to the users' database, containing all the personal information of the registered users~\cite{yahoo-russian-hack}. Although being critical, this episode is not the first (and unfortunately, is not going to be the last) of its kind. Indeed, companies' employees, that have often the privileges to access confidential user information, are usually targeted by external attackers~\cite{facebook-passwords, facebook-compromise, facebook-storing}. 
\\*
History teaches or, at least, it should. Given the enormous data breaches that happened in the past, we expect the Web-sites to improve their security measures, but the reality does not meet the expectations. In our previous work~\cite{raponi2018spark}, we highlighted how almost 44\% of the top 200 Alexa's Web-sites with the domains registered in France, Germany, Italy, Spain, and the U.K., respectively, did manage the recovery of users passwords in a way that could be compromised by malicious users. In this work, the cited Web-sites have been analyzed once again, to verify whether the entry into force of the General Data Protection Regulation (GDPR) has affected their management of the registered users' passwords. As a result, we discovered that only a few Web-sites have strengthened their password management policies, while the vast majority still suffers from the same vulnerabilities. To cope with the above-cited issues, we propose \sol, an innovative solution for password management and recovery that solves some of the critical issues current Web-sites are affected by. Moreover, in this work, we carry out a new analysis of the top 200 Alexa's Web-sites with the domains registered in France, Germany, Italy, Spain, and the U.K. respectively, pointing out that an important percentage of such Web-sites, that are currently among the most visited ones, still suffer from serious privacy concerns and are not compliant with the regulation. Although the impact of these discoveries is significant on its own, the effects on companies are amplified when considering the GDPR, which is violated by the poor security measures adopted by Web-sites. \newline
The GDPR~\cite{gdpr-portal} introduction led to an extensive change in the perception of data security and privacy. It has been proposed as an evolution of the Data Protection Directive (DPP) and has become enforceable since May 25th, 2018. The entry into force of the GDPR imposes enterprises, agencies, organizations, and Web-sites, to guarantee protection by design on the personal data of their users. Failure to comply with this regulation obliges the defaulters to pay substantial fines ranging from 10-20 million euros to a percentage (i.e., 2-4\%) of the previous financial year worldwide turnover---a measure that would likely put out of business quite a few companies, if fined.

In our previous study, one of the most viable and dreadful attacks users were exposed to could be performed by a mail service provider-level attacker, who has the capability to read, modify, and delete the user e-mails, as well as starting the Web-site password recovery mechanism on behalf of the users. To counter this threat, in this paper we propose \sol, an open-source, innovative solution to password recovery that is, by design, resistant to mail-service provider attacks. \sol\ has been released as a module that every Web-site has the opportunity to adopt, allowing registered users to rely on a secure password recovery mechanism. The source code of our proof-of-concept is available here\footnote{In the camera-ready of this manuscript, this will be a hyperlink pointing to the open-source code}. This will allow researchers, practitioners, and industries to check our claims and to benefit of our source code as a ready-to-use basis for their experiments.

\paragraph{Contributions}

In this paper, we provide several contributions: first, we revisit the attacker model, taking into account the password management security of Web-sites; second, we revise the analysis of Web-sites considered in our previous work, showing how most of their administrators have not made any effort to make them compliant with the GDPR; then, we perform a brand new detailed analysis of the current top 200 Alexa's Web-sites with the domains registered in some of the most populated European countries.
Finally, considering that the threat coming from mail service provider-level attackers is real, we propose \sol, an open-source module that could be adopted by any Web-site to provide registered users with a password recovery mechanism preventing mail service provider-level attacks.

\paragraph{Road-map} 

The paper is organized as follows. Section~\ref{sec:related-work} reports the related studies in the literature. Section~\ref{sec:technogical-background} provides a technical background of user authentication on Web-sites and of the password recovery mechanisms they are currently relying upon. The attacker model is described in Section~\ref{sec:adversary-model}, while Section~\ref{sec:Methodology-Results} describes the adopted methodology and presents the results of our analysis of the top Alexa's Web-sites. Section~\ref{sec:maildust} illustrates the implementation of \sol. Finally, Section~\ref{sec:conclusion} draws come concluding remark.

\section{Related Work} \label{sec:related-work}

Although critical, the data breaches suffered by both LinkedIn, in June 2012, and Yahoo, in August 2013, represent only small tiles that make the cyberattacks mosaic up. However, passwords (or rather, their management), which have been the cause of these and many other attacks, are not losing their popularity, still remaining the most adopted authentication mechanism on the web. As soon as they were proposed, the scientific community immediately identified the fragility and the weaknesses of such an authentication technique~\cite{MADHUSUDHAN20121235}. For example, in~\cite{florencio2007large}, the authors reported a large-scale study involving half a million users over three months. A client component installed on the users' machine gathered and extracted both usage and frequency metrics of the passwords typed. This allowed the authors to discover: (i) how many different passwords are used by the users; (ii) how many accounts each user has; (iii) how often passwords are used for more than a web service; and (iv), the type, the length, and the strength of the passwords chosen by the users. According to the study, each user has on average 6.5 passwords, that will be used to access approximately 3.9 different Web-sites. Moreover, each user has 25 accounts requiring passwords and types, on average, 8 passwords per day. As per the strength of the password chosen, every user, if not explicitly asked by the system, choose lowercase-only passwords.

An empirical study about the strength of real-world passwords has been carried out in~\cite{dell2010password}. In this work, the authors implemented state-of-the-art password guessing techniques, including dictionary attacks, Markov chain-based strategies, and mangling relying both on dictionaries and probabilistic context-free grammars, to evaluate their effectiveness over datasets of known passwords. Among their finding, they discovered that users put more effort into choosing the proper username rather than their password. Furthermore, they stressed the importance of the human component when it comes to the robustness of the authentication mechanism. Indeed, if users pay little (or no) attention to the choice of their password, even the most sophisticated and secure systems risk of being compromised.
Given their convenience and the reduced number of false positive and false negative compared to both ``something you are'' and ``something you have'' mechanisms, we expect ``something you know'' mechanisms to be used for years. This makes it crucial to guarantee adequate security for the storage of user's confidential information, as well as secure and protected mechanisms to recover these information in case of theft or forgetfulness. Unfortunately, results in the literature are not encouraging. Indeed, in~\cite{furnell2007assessment}, an assessment of password practices on 10 popular Web-sites have been performed, including Google, Facebook, Yahoo, Amazon, and YouTube. An analysis of the password selection, the password recovery/reset mechanisms, and the restrictions enforced during the password choice found all the Web-sites taken into account inadequate. It is worth noting that the Web-sites considered in the study are among the most visited on the Web, thus making the outcome of the analysis even more worrying. 
A possible solution has been introduced in~\cite{HUANG20111292}, where the authors proposed an authentication service based on one-time passwords, to eliminate the need for preset users passwords during the authentication process. Those passwords are delivered via ubiquitous communication infrastructures like instant messaging services that, as well as mail service providers, should be trusted, a requirement that nowadays cannot be guaranteed.

When a user loses or forgets her password, she generally relies on her e-mail address to recover it, with the strong assumption that she is the only entity having access. In a paper dated back to 2003, S.L. Garfinkel was wondering if the Public Key Infrastructure (PKI), used at that time, could have been substituted by E-mail-based Identification and Authentication (EBIA) methods. The latter consists of using the e-mail address as an identifier and considers the ability to receive e-mails in that address as an authentication mechanism. In the same study, the author identified two main vulnerabilities: (i) the security of this mechanism is strongly dependant on the security of both e-mail servers and passwords; and (ii) the content of e-mails could be accessed by server operators, who can intercept, read, as well as keep copies of e-mail messages directed to the users. This mechanism, although being considered insecure in 2003, is the currently used secondary authentication method nowadays, with the same weaknesses left unresolved.

Authors in~\cite{al2018email} surveyed password recovery e-mails for 50 out of the top English language Web-sites, including Facebook, Dropbox, and Microsoft. The analysis of the password recovery processes has highlighted how, most of them, giving the little attention dedicated to the design, are likely to be extremely prone to compromise. The authors in~\cite{li2018email} realized the criticality of maintaining a single point of failure (the e-mail of the user) as a recovery mechanism. They examined 239 Web-sites, confirming that most of them use e-mails as the standard account recovery mechanism, and built an e-mail-based account recovery attack. After identifying the capabilities of an attacker in this context, they proposed a lightweight e-mail security enhancement against account recovery attacks. The approach adds an extra layer of protection to password recovery e-mails (i.e., e-mails that allow recovering a password). These e-mails will be tagged, and the user can have access only after having successfully passed a 2-Factor Authentication challenge. Although this paper pointed out the vulnerabilities affecting the e-mail-based password recovery methods, the authors did not take into account mail service provider-level adversaries, that would easily bypass the additional security measures adopted. 

Relying on secure e-mail services, such as ProtonMail~\cite{protonmail}, could be an effective countermeasure. ProtonMail, an open-source e-mail service founded by the CERN in 2014, does not log IP addresses and protects the privacy and the anonymity of the users. Moreover, being end-to-end encrypted, the content of the e-mail is indecipherable even by the mail-service provider itself. However, malicious users could still compromise the account of the victim and jeopardize her privacy on-line.

Other alternatives, aiming at improving the security of the secondary authentication mechanism, have been introduced in the literature. In addition to ``something you have'', ``something you are'', and ``something you know'', also ``somebody you know''~\cite{brainard2006fourth} has been taken into account. The authors introduced a process called vouching, in which a helper makes use of her primary authenticator to support an asker in performing the secondary authentication. A prototyping vouching system for SecurID is introduced, where a helper can grant temporary access privileges to an asker who is no longer able to access the service. Although interesting, this method requires a hardware authentication token, hence ``something you have'', a factor that is not implemented by most existing Web-sites. In~\cite{roy2015password}, the authors enriched the password recovery mechanism based on security questions with a keystroke analysis during the user's typing. This allows mitigating brute-force attacks, as well as easily guessable secret questions and answers. However, this approach is renowned for having a high number of false positives, thus jeopardizing the availability of the password recovery mechanism. 

In~\cite{schechter2009s}, the authors introduced a new social authentication approach based on trustees, that will allow users to access their accounts after having forgotten or lost the credentials. Trustees, that are people chosen by the account holders, will have the burden of identifying the account holder either in person (identification by appearance) or by phone (identification by voice). Once the account holder has been identified, she is provided with an account recovery code that will authenticate her in the system. Although this mechanism looks safer if compared with the currently adopted password recovery methods, it may incur in usability problems. Indeed, being people, trustees may not be available at the time of need, making it impossible for the account holder to access the system.

A brief survey on the secondary authentication mechanisms has been introduced in~\cite{reeder2011password}. In this study, every authentication mechanism is evaluated in terms of four criteria: (i) reliability, defined as the likelihood of the account holder's successful authentication; (ii) security, defined as the likelihood that an attacker can impersonate an account holder; (iii) efficiency of the authentication, defined as the time and effort required by the account holder to authenticate; and (iv), efficiency of the setup, defined as the time, the effort, and the privacy sacrificed by both the account holder and the Web-site to configure the authentication mechanism. The secondary authentication mechanisms taken into account include security questions, previously used passwords, e-mail-based verification, printed shared secrets, trustees, in-person proofing, phones, and other services. Although these mechanisms have been studied in detail, the modeling of possible attackers, as well as the study of the security mechanisms adopted by the Web-sites in the wild, were out of the scope of this paper. 

Authentication cookies could be seen as possible substitutes for the user's credentials during the authentication sessions. As for credentials, their disclosure would allow attackers to fully impersonate the user on the Web, exploiting the victim's privileges in the authenticated sessions. In~\cite{calzavara2015supervised}, authors built a dataset gathering 2,464 authentication cookies from a sample of 215 most popular Web-sites of Alexa's ranking. As a result, they proposed the development of a set of binary classifiers, aimed at identifying these authentication cookies exploiting (supervised) machine learning techniques. 

Considering the vulnerabilities identified in the mechanisms currently in use, the need for finding alternatives is evident, to provide users with adequate security and privacy on the Web.

\section{Background}
\label{sec:technogical-background}
In this section, the technical background of user authentication mechanisms is provided, as well as a description of the password recovery mechanisms Web-sites are currently relying upon. Furthermore, the Shamir's Secret Sharing scheme is introduced, instrumental to the realization of \sol.

\subsection{User authentication on Web-sites}

Authenticating on a Web-site refers to the process in which an account holder provides her credentials to the service. This service, once received the credentials, compares them with those stored in the Web-site database (or in a cloud server) to check whether there is a match. If such a match is found, the authentication process is successful and the requesting account holder is granted authorization to access. Several authentication methods could be implemented by a Web-site~\cite{furnell2007comparison}, and a combination of them leads to more accurate identification of the user.

\begin{itemize}
	\item[-] \textbf{1-Factor-Authentication (1FA)}: In this authentication mechanism, only one factor (e.g., ``something the user knows'') is required to authenticate a user. Being the simplest and the less expensive, 1FA is the most common authentication mechanism adopted by Web-sites;
	\item[-] \textbf{2-Factor-Authentication (2FA)}: In this authentication mechanism, two factors are required to authenticate a user. In addition to ``something the user knows'', the user has to present either ``something the user has'' or ``something the user is''. Examples of ``something the user has'' include smart cards or physical token, while ``something the user is'' refers to biometrical information of the user, such as her voice, hand geometry, retinal, iris, fingerprints, and possibly others. Many of the sensitive Web-sites (e.g., banks) adopt this mechanism to strongly authenticate their users, while it remains optional in others, such as Gmail~\cite{google-2fa} and ProtonMail~\cite{protonmail-2fa}.
	\item[-] \textbf{3-Factor-Authentication (3FA)}: In this authentication mechanism, the user needs to be authenticated with ``something she knows'', ``something she has'', and ``something she is''. Although it is the most secure of the three authentication mechanisms, 3FA is also the most expensive one and, therefore, has not found application on Web-sites yet.
\end{itemize}

\subsection{Password recovery}
Password recovery mechanisms, implemented by Web-sites, allow registered account holders to recover their secret passwords in case of theft or forgetfulness. There are many types of password recovery mechanisms, each with its characteristics and weaknesses. The mechanisms that have been frequently adopted by the Web-sites are reported below.

\subsubsection{Security Questions}
In this password recovery mechanism, the requesting user is asked security questions that, if answered correctly, allow the account holder to access. \\ \\
\textit{Assumption.} Only the account holder is able to correctly answer the security questions. \\ \\
The security questions could be standard, such as ``What is your mother's maiden name?'' or ``What is the name of your cat?'', or could be customized by the users. Regardless of this choice, many studies in the literature demonstrated the limitations and the vulnerabilities of such a password recovery mechanism. Indeed, in the information era, using a set of standard security questions is quite insecure. Finding answers to questions such as ``What is the name of your primary school?'' or ``Who is your favorite actor?'' becomes trivial by surfing the personal information the account holder probably shared on social networks. Even without relying on a standard set of security questions, but on a customized set of them, several weaknesses have been identified. Indeed, according to~\cite{just2009personal, reeder2011password}, users should select security questions that are memorable, reasonably unpopular with other users, not researchable on-line, and unknown by any untrusted acquaintances.

\subsubsection{Previously used passwords}
In this password recovery mechanism, to get access to the account, the user will be asked to provide one (or a set of) passwords that she has previously used in the same Web-site. \\ \\
\textit{Assumption} The account holder, and only her, remembers all the passwords she previously used on the Web-site. \\ \\

Given that each person is able to remember only a limited number of passwords, this password recovery mechanism inevitably leads to the use of the same passwords for more than one Web-site. Adopting this mechanism, although with some limitations (e.g., the account holder may not remember the passwords she previously used in that specific Web-site) and some security problems (e.g., a malicious user may have gotten some of the passwords the account holder previously used in that Web-site) is safer if compared with the password recovery mechanisms currently adopted by Web-sites. It is worth to point out that this mechanism cannot be adopted if the first password has been lost or stolen (i.e., there are no previously used passwords to rely upon).

\subsubsection{E-mail-based authentication}
The password recovery mechanism that is predominantly used by Web-sites is the e-mail-based password recovery mechanism in which an account holder, having lost or forgotten her credentials, relies on her e-mail account to get access to the Web-site~\cite{garfinkel2003email, reeder2011password}. \\ \\
\textit{Assumption.} The account holder is the only entity that has access to her e-mail account. \\ \\
The different ways to implement the e-mail password recovery mechanism, together with the possible related security vulnerabilities, are detailed in the following.

\begin{itemize}
    \item[-] \textit{old password by e-mail} (\textit{Old Pw}).
    Once the account holder starts the password recovery procedure, the Web-site sends her an e-mail containing her original password. This is the most dangerous way to implement the password recovery mechanism. Indeed, by knowing the original password of the account holder, the Web-site proves not to use hash functions or other mechanisms to avoid storing the passwords of the registered users in clear. Given that most of the users make use of a limited number of passwords to access many different Web-sites~\cite{reusing-password}, this behavior could bring to catastrophic consequences. If an attacker gets his hands on the password database of the Web-site, both the security and the privacy of all the registered users are jeopardized. It is worth to notice that this way of implementing the e-mail-based password recovery mechanism has been deprecated more than 30 years ago~\cite{fighting-computer-crime}.
    \item[-] \textit{sending a new temporary password by e-mail} (\textit{Temp Pw});
    Once the account holder starts the password recovery procedure, the Web-site sends her an e-mail containing a temporary password, that she is forced to change at the first access.
	\item[-] \textit{sending a new (not temporary mail} (\textit{New Pw});
	Once the account holder starts the password recovery procedure, the Web-site sends her an e-mail containing a new password. The Web-site does not force the account holder to change the password at the first access. In this paper we consider: \textit{weak} the passwords with less than $2^{50}$ combinations; \textit{medium} the passwords with more than $2^{50}$ combinations; and \textit{strong} the passwords with more than $2^{70}$ combinations; respectively.
	\item[-] \textit{sending an HTTP link by e-mail} (\textit{HTTP link});
	Once the account holder starts the password recovery procedure, the Web-site sends her an e-mail containing an HTTP link. If clicked, this link redirects the account holder to the Web-site to choose a new password. In this study, we consider ``vulnerable'' the Web-sites that implemented the e-mail password recovery mechanism this way. Indeed, the HTTP protocol does not guarantee any insurance with respect to both man-in-the-middle and snooping attacks. Being not encrypted, the communication between the account holder's browser and the Web-site could be intercepted, eavesdropped, and modified, thus jeopardizing both the integrity and the confidentiality of the communication.
	\item[-] \textit{sending an HTTPS link by e-mail} (\textit{HTTPS link}).
	Once the account holder starts the password recovery procedure, the Web-site sends her an e-mail containing an HTTPS link. If clicked, this link redirects the account holder to the Web-site to choose a new password. It is worth to notice that this is the safest way of implementing the e-mail-based password recovery procedure.
\end{itemize}

\subsubsection{Other password recovery mechanisms}
The literature boasts many studies proposing alternative approaches to efficiently implement password recovery on Web-sites. Some of them propose a kind of multi-device authentication, where a cell-phone receives an SMS message (or a phone call) from the Web-site, containing information to access the account. As for e-mail-based authentication, also this approach relies on the assumption that only the device holder is able to access a secret sent to the device~\cite{reeder2011password}. Multiple devices are also needed for the solution proposed by~\cite{brainard2006fourth}, where the authors introduced a process called vouching, in which a helper makes use of her primary authenticator to support an asker in performing the secondary authentication. A prototyping vouching system for SecurID is introduced, where a helper can grant temporary access privileges to an asker who is no longer able to access the service. This approach is safer than the ones currently implemented but requires a hardware authentication token. Other approaches, instead, rely on people to securely recover the password. In~\cite{schechter2009s}, for example, the authors introduced a new social authentication approach based on trustees. Trustees are people selected by the account holder that will have the burden of identifying her either in person (identification by appearance) or by phone (identification by voice). Once the account holder has been identified, she is provided with an account recovery code that will authenticate her in the system. As highlighted in the previous section, this mechanism is safer if compared with the currently adopted passwords recovery mechanisms but it may incur in availability problems, that would jeopardize its usability.

\subsection{Shamir's Secret Sharing Scheme}
The Shamir's Secret Sharing Scheme~\cite{shamir1979share} allows to divide any data D into n tokens $D_1 \dots D_n$ , such that:
\begin{itemize}
    \item[-] $k$ tokens ($k \le N$) will be enough to reconstruct D; and,
    \item[-] the knowledge of $k-1$ tokens reveals absolutely no information about D.
\end{itemize}
The Shamir's Secret Sharing Scheme is based on polynomial interpolation: in a bi-dimensional plane, given j points $(x_1, y_1), \dots, (x_j, y_j)$ with $x_1 \ne x_2 \ne \dots \ne x_j$ there is only one polynomial function $f$ of degree $j-1$ such that $\forall i$ $f(x_i) = y_i$. \newline
Given this premise, to divide D into tokens $D_1 \dots D_n$, a random $j-1$ degree polynomial $f(x) = t_0 + t_1x + \dots + t_{j-1}x^{j-1}$ is picked, where $t_0 = D$. Then the function $D_1 = f(1), D_2 = f(2), \dots , D_i = f(i), \dots , D_n = f(n)$ is evaluated. Any k-sized subset of these $D_i$ values and their identifying indices allows to find the coefficients of $f(x)$ by interpolation, thus evaluating $D = f(0)$. \newline
We apply the Shamir's Secret Sharing Scheme to the password generated by our open-source password recovery module. The tokens obtained will be distributed over the password recovery resources provided by the user during the registration to the Web-site. The goal is to allow users to reconstruct the new password \textcolor{black}{even if} one or more password recovery resources \textcolor{black}{are} compromised.

\section{Adversary Model} \label{sec:adversary-model}

Given the dynamic ecosystem of Web-sites, we model several categories of adversaries, summarized in~\autoref{tab:attackers-description}. An adversary may be \textit{passive} or \textit{active}. A \textit{passive} attacker does not interact with the Web-site, while an \textit{active} attacker could perform different actions, including starting the password recovery procedure on behalf of the account holder. Furthermore, an adversary may be \textit{detectable} or \textit{undetectable}. An attacker is detectable if, once impersonating the account holder, is forced to leave traces (i.e., the account holder has a chance of being aware, or at least suspicious, that an impersonation could have happened). Conversely, an attacker is \textit{undetectable} if she has the opportunity of impersonating the account holder without leaving any trace. 

The goal of the attacker, being her active or passive, is to obtain the credentials of the account holder to impersonate the victim on the Web-site. Being detectable or undetectable strongly depends on the password recovery mechanism adopted by the Web-site.

We introduce four possible attacks against a target account holder, who has an account on a target Web-site:

\begin{itemize}
	\item[-] Mail service provider-level attack;
	\item[-] Web server intruder attack;
	\item[-] Client intruder attack; and,
	\item[-] Sniffing attack.
\end{itemize}

\begin{table*}[h]
	\caption{Attackers types}
	\label{tab:attackers-description}
	\centering
    \begin{adjustbox}{max width=\textwidth}	
    \begin{tabular}{|C{2.4cm}|C{4.9cm}|C{4.9cm}|}
	    \hline
		\textbf{Type} & \textbf{Passive} & \textbf{Active} \\
		\hline
		\textbf{Detectable} & No ability of interacting with the Web-site and traces will be left behind & Ability of interacting with the Web-site but traces will be left behind \\
		\hline
		\textbf{Undetectable} & No ability of interacting with the Web-site but no traces will be left behind & Ability of interacting with the Web-site but no traces will be left behind.\\
	\hline
	\end{tabular}
	\end{adjustbox}
\end{table*}

\begin{table*}[h]
	\caption{Attackers resources and possible accesses}
	\label{tab:attackers-possible-accesses}
	\centering
	\begin{adjustbox}{max width=\textwidth}
	\begin{tabular}{|C{3.8cm}|C{2.2cm}|C{3cm}|C{3cm}|}
	    \hline
		\textbf{Attackers/Access} & \textbf{User e-mails} & \textbf{Web-site password DB} & \textbf{Web-site password recovery method} \\
		\hline
		Mail service provider-level & \cmark & \xmark & \cmark \\
		\hline
		Web server intruder & \xmark & \cmark & \cmark \\
		\hline
		Client intruder & \cmark & \xmark & \cmark \\
		\hline
		Sniffing & \xmark & \xmark & \cmark \\
		\hline
	\end{tabular}
	\end{adjustbox}
\end{table*}

A \textit{mail service provider-level attack} can be performed by a malicious service provider, as well as by a malicious user that compromised the mail service provider. In this case, the attacker would have access to every e-mail of the account holder, thus having the opportunity to obtain a lot of sensitive and confidential information. A \textit{web server intruder attack} can be performed by an attacker that violated the Web-site. Such an attacker would have access to the database in which the passwords of the users are stored. A \textit{client intruder attack} can be performed by an adversary that violated a user's device. This adversary may have stolen the device or obtained some kind of remote access on it. Finally, the \textit{sniffing attack} can be performed by any attacker that has information about the Web-site password recovery method. It is worth to notice that obtaining such information is easy, since every user can register an account and pretend to have lost the password. Furthermore, this last attacker has the opportunity to sniff the traffic between the client and the Web-site during their communication. Without loss of generality, we assume that the sniffer attacker cannot read the content of the e-mails.

The resources the adversaries can access and the actions the adversaries can perform are summarized in~\autoref{tab:attackers-possible-accesses}, while their capabilities are described in Section~\ref{sec:attackers-capabilities}.

\subsection{Attackers capabilities} \label{sec:attackers-capabilities}
The capabilities of the attackers, as well as their characteristics, are described in this section. The result of the analysis are summarized in~\autoref{tab:passive-attackers-synoptic-table} for passive attackers, and in~\autoref{tab:active-attackers-synoptic-table} for active attackers.

\begin{table*}[h]
	\caption{Synoptic table related to passive attackers}
	\label{tab:passive-attackers-synoptic-table}
	\centering
	\begin{adjustbox}{max width=\textwidth}
	\begin{tabular}{|C{1.95cm}|C{1.95cm}|C{1.95cm}|C{1.95cm}|C{1.95cm}|C{1.95cm}|}
	    \hline
		\textbf{Attacks / Recovery methods} & \textbf{Old Pw} & \textbf{New Pw} & \textbf{Temp Pw} & \textbf{HTTP link} & \textbf{HTTPS link} \\
		\hline
		Mail service provider-level & \textit{undetectable} & \textit{undetectable} & \textit{detectable} & \textit{detectable} & \textit{detectable} \\
		\hline
		Web server intruder & \textit{undetectable} & \textit{storage method} & \textit{storage method} & \textit{storage method} & \textit{storage method} \\ 
		\hline
		Client intruder & \textit{undetectable} & \textit{undetectable} & \textit{detectable} / \textit{user's behavior} & \textit{detectable} / \textit{user's behavior} & \textit{detectable} / \textit{user's behavior} \\
		\hline
		Sniffing & \textit{undetectable} & \textit{undetectable} & \textit{undetectable} & \textit{undetectable} & \textit{undetectable} \\
		\hline
	\end{tabular}
	\end{adjustbox}
\end{table*}

\subsubsection{Mail service provider-level attacker}
By having access to the e-mails of the account holder (i.e., the emergency authentication mechanism adopted by most Web-sites), a \textit{passive mail service provider-level attacker} may obtain the credentials of the victim in any case, regardless of which password recovery mechanism the Web-site adopts.
However, in case the Web-site adopts either \textit{Old Pw} or \textit{New Pw} recovery mechanisms, the attacker would be able to impersonate the account holder without leaving any trace behind, thus remaining undetectable. Indeed, the attacker could read the original password sent by e-mail to the account holder by the Web-site. From that moment on, the account holder and the attacker could share the same account on the Web-site, without the former being aware of it. The attacker would be detectable in case the Web-site adopts any other e-mail-based password recovery mechanism. For example, in case the Web-site adopts \textit{HTTP link}, \textit{HTTPS link}, or \textit{Temp Pw}, the attacker could access the link on behalf of the account holder, thus having access to the Web-site, but will be forced to change the password, no longer granting access to the account holder. However, The attacker may delete the e-mail after having accessed the link (or used the new password). When the account holder wants to access the system her password will not work, but she could think of a malfunction of the Web-site. In the meantime, the adversary may have logged-in, obtained the needed information, and logged-out, with good chances of not being suspected at all. \newline
If the Web-sites stores the passwords of the users in clear, also the \textit{active mail service provider-level attacker} could remain undetectable. Indeed, when the attacker starts the recovery procedure on behalf of the account holder, she can obtain the original password by-email. The e-mail could then be deleted before the victim could access it. In the event other password recovery mechanisms are adopted, the password would change. Even in this case, the compromising e-mail could be deleted by the mail-service provider before the victim has the opportunity of accessing it. The account holder would have no longer access to the Web-site, but she could wrongfully think of a Web-site malfunction or to have forgotten the password, thus exonerating the adversary.

\subsubsection{Web server intruder attacker}

The password storage management of the (violated) Web-site is determining to understand which information a \textit{passive Web server intruder attacker} may obtain. Indeed, if the Web-site stores the passwords in clear (i.e., without relying on hash functions), the attacker could transparently make use of the credentials of the victim to get access. \\
An \textit{active Web server intruder attacker} would have the same capabilities of a passive one, dictated by the password storage management adopted by the Web-site. The only difference concerns the traces left behind. Indeed, in case the attacker interacts with the password recovery module of the Web-site on behalf of the victim, she will not be able to delete the e-mail. The account holder would receive an e-mail containing instructions for recovering the password and would become suspicious.

\begin{table*}[h]
	\caption{Synoptic table related to active attackers}
	\label{tab:active-attackers-synoptic-table}
	\centering
	\begin{adjustbox}{max width=\textwidth}
	\begin{tabular}{|C{1.95cm}|C{1.95cm}|C{1.95cm}|C{1.95cm}|C{1.95cm}|C{1.95cm}|}
		\hline
		\textbf{Attacks / Recovery methods} & \textbf{Old Pw} & \textbf{New Pw} & \textbf{Temp Pw} & \textbf{HTTP link} & \textbf{HTTPS link} \\
		\hline
	Mail service provider-level & \textit{undetectable} & \textit{detectable} & \textit{detectable} & \textit{detectable} & \textit{detectable} \\
		\hline
		Web server intruder & \textit{undetectable} & \textit{detectable} / \textit{storage method} & \textit{detectable} / \textit{storage method} & \textit{detectable} / \textit{storage method} & \textit{detectable} / \textit{storage method} \\ 
		\hline
		Client intruder & \textit{undetectable} & \textit{detectable} & \textit{detectable} & \textit{detectable} & \textit{detectable} \\
		\hline
		Sniffing & \textit{detectable - useless} & \textit{detectable - easier} & \textit{detectable - easier} & \textit{detectable} & \textit{detectable} \\
		\hline
	\end{tabular}
	\end{adjustbox}
\end{table*}

\subsubsection{Client intruder attacker}
Most likely having access to the e-mails of the account holder (i.e., most of the users do not want to insert their passwords from scratch every single time, allowing the browser to remember them~\cite{password-life-cycle}), a \textit{passive client intruder attacker} that has access to the device of the account holder would have the same capabilities of a mail service provider level attacker. In case the account holder does not rely on browsers shortcuts to remember the credentials, an attacker with either remote or temporary physical access to the device has the opportunity to install a key-logger software. This would allow the attacker to capture the input of the account holder, including the password typed. Besides, the remote (or physical) access to the device would allow the attacker to find password files of the account holder~\cite{password-memorability}. In this case, the attacker's capabilities will be dependant on the behavior (and the habits) of the account holder. \\
The capabilities of an \textit{active client intruder attacker} are the same of the passive client intruder attacker, since the interaction with the Web-site would not bring any benefit.

\subsubsection{Sniffing attacker}
A \textit{passive sniffing attacker} has mainly three ways to obtain the credentials of an account holder: (i) the attacker intercept the HTTP communications between the account holder and the Web-site; (ii), the attacker performs a man-in-the-middle attack during an HTTPS communication between the account holder and the Web-site; or (iii), the adversary brute-forces the Web-site to guess the credentials of the account holder, respectively. In the brute-forcing event, in case the Web-site does implement neither protection mechanisms (e.g., Completely Automated Public Turing test to tell Computers and Humans Apart (CAPTCHA), blocking the access after a maximum number of login attempts) nor warning systems (e.g., e-mail or SMS to the account holder after a maximum number of login attempts), if the attacker guesses the password she will be able to impersonate the victim before the latter being aware of it. \\
In other cases, the capabilities of an \textit{active sniffing attacker} depend on the password recovery mechanism adopted by the Web-site: 
\begin{itemize}
    \item \textit{Old Pw}: when the attacker starts the password recovery procedure on behalf of the user, the Web-site sends the account holder an e-mail containing her password. This procedure, besides making the attacker detectable (because the user will receive the e-mail) is also fruitless, since the attacker would not get any advantage (i.e., the password to brute-force would be the same).
    \item \textit{New Pw} or \textit{Temp Pw}: if the Web-site adopts such password recovery mechanisms, the password recovery procedure started by the attacker on behalf of the account holder would give her precious information about the password to brute-force. Indeed, even if the user during the registration chose an incredibly strong password, the attacker would ask for a new one on behalf of the account holder. This gives the attacker useful information about the password to brute-force, since she would know both its structure and its level of security. Let us suppose that the account holder has an extremely hard-to-guess password. The attacker, that wants to impersonate the account holder, creates a personal account on the target Web-site and asks (for several times) for the password recovery procedure, thus obtaining information about the passwords generated by the Web-site (e.g., 6-characters passwords without capital letters). Once obtained this information, the attacker would start the recovery procedure on behalf of the account holder that, without being aware of it, will be associated with an easier-to-guess password.
    \item \textit{HTTP link or HTTPS link}: without having access to the e-mails of the user, the attacker will become detectable and will not gain any additional information about the password.
\end{itemize}

Note that being active or passive has consequences on both the \textit{attack timing} and the \textit{attack extension}. Indeed, on the one hand, active attackers may obtain the credential of the account holder at any time, while passive attackers have to wait for an action from the account holder. On the other hand, an active attacker has to be controlled by a human, to take immediate actions, while passive attackers may be implemented as autonomous software and triggered when specific events happen (e.g., receiving an e-mail with links or credentials). This makes passive attackers easy to implement and to spread.

\section{Methodology and Results} \label{sec:Methodology-Results}
In this section, we first describe the methodology to analyze the security of the Web-sites' password management. Then, we provide the results of three analysis: in the first analysis we analyze the password management security of the top Alexa's 200 Web-sites with the domains registered in five European countries: France, Germany, Italy, Spain, and the U.K., respectively, in the period of December 2017; in the second analysis we analyze the same Web-sites of the same countries in the period of December 2018 (i.e., one year later); and, finally, we perform the current top 200 Alexa's Web-sites of the same countries in the period of December 2018. The goal is to understand whether the entry into force of the GDPR has affected the password management of both the Web-sites that were popular (i.e., the most visited ones) at the time of the first analysis (i.e., December 2017) and the Web-sites that are popular one year later (i.e., after the GDPR's entry into force).

\subsection{Methodology} \label{sec:methodology}
In the methodology we made two independent choices: (i) how to choose the Web-site to consider for the analysis; and (ii) how to analyze the selected Web-sites. As for the first choice, we relied on the Amazon Alexa Web service~\cite{amazon-alexa}. Amazon Alexa provides a Web service called ``top sites'', in which Web-sites are listed according to the Alexa's Traffic Rank. Such a ranking, that is daily updated, is determined by a combination of unique visitors and page views of Web-sites~\cite{amazon-alexa-traffic-ranking}. We first selected a subset of European countries subject to the GDPR regulations: France, Germany, Italy, Spain, and the U.K.--till Brexit happens, the U.K. is subject to GDPR as well, respectively. Then, for each country, we took into account the first 200 Web-sites according to the ranking of Alexa's top sites. We chose the top ones since, as the ranking goes down, the impact of the Web-sites (that are less and less visited) decreases. We created an account on the Alexa's Web-sites and we started the collection of the URLs.
It is worth to notice that Amazon Alexa provides multiple ways to obtain the top-sites. Indeed, they may be grouped according to three major filters: global, by country, and by category. As the name suggests, the global division allows to obtain the most visited Web-sites globally. The ``by country'' division, instead, allows to obtain the most visited Web-sites from the selected countries. The domain of these Web-sites could not be registered in the selected countries (e.g., YouTube.com, the most famous video-sharing platform, although being registered in California, United States, is in the top five of all the countries we considered in this study). The third division allows to obtain the top Web-sites according to further labels (e.g., arts, regional, sports, computers to name a few). With this filter we were able to choose the specific country, thus obtaining the top 200 Web-sites with the domains registered in the countries we selected. Once obtained the URLs of the Web-sites, we performed a detailed analysis to study the security of their password storage management.

The second phase starts with the selection of the Web-sites requiring user registration among the top 200 of each country. Once selected, a further filtering allowed us to remove the Web-sites that, during the registration, required the user to insert privileged information (e.g., the account number for banks, the customer code for wholesalers the id numbers for universities, and possibly others). After this filtering phase, we manually analyzed each Web-site to obtain detailed information about the password management. It is worth to notice that, although this activity took some amount of time (i.e., around three months), the quality and the detail of the results are invaluable.
In detail, for each of the Web-site to analyze:

\begin{itemize}
	\item[-] we registered a user account with a new e-mail gdpr.experiment@gmail.com, a Gmail account created ad-hoc for the analysis;
	\item[-] we started the recovery procedure, pretending to have lost the password; and,
	\item[-] we collected information about the password recovery mechanism.
\end{itemize}

\subsection{First Analysis: Top 200 Alexa's Web-sites with the domains registered in five European countries (December 2017)} \label{sec:first-analysis}

In this section, we provide the analysis of the top 200 Alexa's Web-sites with the domains registered in five European countries: France, Germany, Italy, Spain, and the U.K., respectively, in the period of December 2017. We first provide details about the filtering phase, in which we remove Web-sites we are not able to analyze, then we introduce the password recovery mechanisms adopted by the Web-sites we take into account.

\subsubsection{Filtering phase}
As described in Subsection~\ref{sec:methodology}, the Web-sites have been filtered to remove both those that do not require authentication and the ones requiring, for the registration phase, information that we are not able to produce (e.g., id numbers for universities, account number for banks) Of the 1,000 Web-sites, 722 did not require authentication---and hence were excluded by our study---, while out of the remaining 278, we focused on 174---since 104 demanded information we could not produce. The results of this filtering are shown in~\autoref{tab:analyzed-websites}: most of the Web-sites taken into account have the domain registered in the U.K., (i.e., 71, approximately 40.8\%), while only a few Web-sites from Germany and Spain appear to require authentication from the user, 17 ($\approx$ 9.7\%) and 19 ($\approx$ 10.9\%), respectively.

\begin{table}[h]
    \small
	\caption{Number of analyzed Web-sites among the top 200 per country (December 2017)}
	\label{tab:analyzed-websites}
	\centering
	\begin{adjustbox}{max width=\textwidth}
	\begin{tabular}{|C{3cm}|C{3cm}|}
	    \hline
		\textbf{Country} & \textbf{Web-sites (\#)} \\
		\hline
		France & 31\\
		\hline
		Germany & 19\\
		\hline
		Italy & 36\\
		\hline
		Spain & 17\\
		\hline
		U.K. & 71\\
		\hline
		\textbf{Total} & \textbf{174}\\
		\hline
	\end{tabular}
	\end{adjustbox}
\end{table}

\begin{table*}[h]
	\caption{Password recovery mechanisms adopted by the analyzed Web-sites (December 2017)}
	\label{tab:analysis-result}
	\centering
	\begin{adjustbox}{max width=\textwidth}
	\begin{tabular}{|C{1.4cm}C{1.5cm}|C{1.5cm}C{1.9cm}C{1.5cm}C{1.6cm}|C{1.5cm}}
	    \hline
		\multirow{2}{*}{\textbf{Country}} & \textbf{Web-sites (\#)}  & \textbf{Old Pw}  & \textbf{New Pw} & \textbf{Temp Pw} & \textbf{HTTP link} & \multicolumn{1}{C{2cm}|}{\textbf{Vulnerable Web-sites (\%)}} \\
		\hline
		France & 31 & 0 & 10 & 0 & 7 & \multicolumn{1}{c|}{54.84}\\
		\hline
		Germany & 19 & 1 & 1 & 0 & 5 & \multicolumn{1}{c|}{36.84}\\
		\hline
		Italy & 36 & 4 & 11 & 1 & 4 & \multicolumn{1}{c|}{55.55}\\
		\hline
		Spain & 17 & 2 & 5 & 0 & 1 & \multicolumn{1}{c|}{47.06}\\
		\hline
		U.K. & 71 & 1 & 5 & 2 & 16 & \multicolumn{1}{c|}{33.8} \\
		\cline{1-7}
		\textbf{Total} & \textbf{174} & \textbf{8 (4.6\%)} & \textbf{32 (18.4\%)} & \textbf{3 (1.7\%)} & \textbf{33 (19\%)} & \\
		\cline{1-6}
	\end{tabular}
	\end{adjustbox}
\end{table*}

The results of the analysis are shown in~\autoref{tab:analysis-result}. Each column of the table represents a different \textcolor{black}{password recovery mechanism} adopted by the analyzed Web-sites---these password recovery mechanisms can be observed from left to right in decreasing order of severity. Of the 174 Web-sites, 8 Web-sites \mbox{($\approx$ 4.6\%)} make use of the \textit{Old Pw} password recovery mechanism, thus storing the passwords in clear, 32 Web-sites ($\approx$ 18.4\%) make use of the \textit{New Pw} password recovery mechanism, 3 Web-sites ($\approx$ 1.7\%) use the \textit{Temp Pw} password recovery mechanism, while 33 Web-sites ($\approx$ 19\%) allow users to recover the password through an \textit{HTTP link}, for a total of 43.7\% of vulnerable web-sites.

\begin{table*}[h]
	\caption{Password robustness analysis}
	\label{tab:password-analysis}
	\centering
	\begin{adjustbox}{max width=\textwidth}
	\begin{tabular}{|C{1.8cm}|C{1.28cm}|C{1.28cm}|C{1.28cm}|C{1.28cm}|C{1.28cm}|C{1.28cm}|C{1.28cm}|}
		\hline
		\textbf{Password} & $>2^{10}$ & $>2^{20}$ & $>2^{30}$ & $>2^{40}$ & $>2^{50}$ & $>2^{60}$ & $>2^{70}$ \\
		\hline
		New & 100\% & 93.75\% & 75\% & 62.5\% & 3.125\% & 3.125\% & 3.125\% \\
		\hline
		Temporary & 100\% & 100\% & 100\% & 100\% & 33.33\% & 0 & 0 \\
		\hline
	\end{tabular}
	\end{adjustbox}
\end{table*}

Given the popularity of the \textit{New Pw} password recovery mechanisms, we have carried out further investigations to analyze the security behind the process of generating new passwords. In particular, we obtained this information by requesting a new password 125 times, and by analyzing the passwords provided by the Web-sites. For example, \textit{AKRBA2Y1}, \textit{321fr26f}, \textit{xlpeor} are three passwords sent by a given Web-site after having started the recovery procedure. In this case, the passwords are composed of uppercase or lowercase letters, number, but no special characters, with an overall length of maximum 8 characters. The strength of the passwords created by this Web-site has been computed as $62^8 \approx 2^{48}$.~\autoref{tab:password-analysis} details the analysis of the passwords; according to our evaluation, only 3.125\% of Web-sites choose passwords that could be considered decent, while more than 90\% are considered weak. 
This poor password choice makes the Web-sites prone to brute-force attacks performed by sniffing attackers.

\subsection{Second Analysis: Top 200 Alexa's Web-sites with the domains registered in the five European countries---one year later (in December 2018)}
In this section, we provide a new analysis of the same 200 Alexa's Web-sites of the five countries carried out in December 2017, one year later (i.e., in December 2018). The goal is to understand if the entry into force of the GDPR has \textcolor{black}{led to improvements on} the Web-site's password management security.

\begin{table*}[]
    \centering
    \caption{Old Analysis (December 2017) vs New Analysis (December 2018)}
    \label{tab:old-analysis-vs-new-analysis}
    \begin{adjustbox}{max width=\textwidth}
    \begin{tabular}{|c|cc|cc|cc|cc|cc|cc}
        \hline
        \multirow{3}{*}{\textbf{Country}} & \multicolumn{2}{C{1.5cm}|}{\textbf{Web-Sites (\#)}} & \multicolumn{2}{c|}{\textbf{Old Pw}} & \multicolumn{2}{c|}{\textbf{New Pw}} & \multicolumn{2}{c|}{\textbf{Temp Pw}} & \multicolumn{2}{C{1.5cm}|}{\textbf{HTTP Link}} & \multicolumn{2}{C{2cm}|}{\textbf{Vulnerable Web-sites (\%)}} \\ 
        & '17 & '18 & '17 & '18 & '17 & '18 & '17 & '18 & '17 & '18 & '17 & \multicolumn{1}{c|}{'18}\\
        \hline
        France & 31 & 33 & 0 & 0 & 10 & 10 & 0 & 0 & 7 & 5 & 54.84 & \multicolumn{1}{c|}{45.45} \\
        Germany & 19 & 16 & 1 & 1 & 1 & 1 & 0 & 0 & 5 & 4 & 36.84 & \multicolumn{1}{c|}{37.5} \\
        Italy & 36 & 34 & 4 & 4 & 11 & 11 & 1 & 1 & 4 & 3 & 55.55 & \multicolumn{1}{c|}{55.88} \\
        Spain & 17 & 17 & 2 & 2 & 5 & 4 & 0 & 0 & 1 & 2 & 47.06 & \multicolumn{1}{c|}{47.06} \\
        U.K. & 71 & 74 & 1 & 1 & 5 & 3 & 2 & 2 & 16 & 16 & 33.8 & \multicolumn{1}{c|}{29.7} \\
        \hline
        \textbf{Total} & \textbf{174} & \textbf{174} & \textbf{8} & \textbf{8} & \textbf{32} & \textbf{29} & \textbf{3} & \textbf{3} & \textbf{33} & \textbf{30} & & \\
        \cline{1-11}
    \end{tabular}
    \end{adjustbox}
\end{table*}

\autoref{tab:old-analysis-vs-new-analysis} depicts the changes that have taken place in the Web-sites' password recovery management one year apart---before and after the entry into force of the GDPR, respectively. Note that, in spite of some negligible changes, the Web-sites did not do their best to become compliant with the regulation. Indeed, in December 2017, 43.7\% of Web-sites were vulnerable, while in December 2018 the percentage became 43.12, showing an almost negligible improvement. In the following, for each country, we specify what changes have been highlighted by the analysis.

\subsubsection{United Kingdom}
The English Web-sites have imperceptibly improved the security of their password recovery management, indeed the percentage of the vulnerable Web-sites decreases from 33.8\% to 29.7\%. In detail, by the results of the new analysis, we note that:
\begin{itemize}
    \item[-] 68 Web-sites have not changed their password recovery mechanism;
    \item[-] 4 Web-sites that were not working during the first analysis now work, and allow the registration of user accounts. The password recovery mechanisms adopted are \textit{HTTPS link} for 3 Web-sites and \textit{HTTP link} for 1 Web-site, respectively;
    \item[-] 2 Web-sites have changed their password recovery mechanism from \textit{New Pw} to \textit{HTTPS link}; and
    \item[-] 1 Web-site (that was managing its password recovery with the \textit{HTTP link} mechanism) no longer allows the user registration.
\end{itemize}

\subsubsection{France} French Web-sites have improved the security of their password recovery management more significantly, lowering the percentage of vulnerable Web-sites from 54.84\% to 45.45\%. In detail, by the results of the new analysis, we note that:
\begin{itemize}
    \item[-] 27 Web-sites have not changed their password recovery mechanism;
    \item[-] 2 Web-sites that were not working during the first analysis now work, and allow the registration of user accounts. These Web-sites both adopt the \textit{HTTPS link} password recovery mechanism;
    \item[-] 3 Web-sites have changed their password recovery mechanism from \textit{HTTP link} to \textit{HTTPS link}; and
    \item[-] 1 Web-site has changed its password recovery mechanism from \textit{HTTPS link} to \textit{HTTP link}.
\end{itemize}

\subsubsection{Germany} The German Web-sites have not made security improvements for password recovery management, in fact, the percentage of vulnerable Web-sites has changed from 36.84\% to 37.5\%. In detail, by the results of the new analysis, we note that:
\begin{itemize}
    \item[-] 15 Web-sites have not changed their password recovery mechanism;
    \item[-] 1 Web-site has changed its password recovery mechanism from \textit{HTTP link} to \textit{HTTPS link};
    \item[-] 2 Web-sites do not work anymore, they were both employing the \textit{HTTPS link} mechanism; and
    \item[-] 1 Web-site (that was managing its password recovery with the \textit{HTTPS link} mechanism) no longer allows the user registration.
\end{itemize}

\subsubsection{Italy} Most of the Italian Web-sites, as well as the German ones, have not made efforts to be compliant with the regulation. The percentage of vulnerable Web-sites has changed from 55.55\% to 55.88\%. In detail, by the results of the new analysis, we note that:
\begin{itemize}
    \item[-] 31 Web-sites have not changed their password recovery mechanism;
    \item[-] 1 Web-site has changed its password recovery mechanism from \textit{HTTP link} to \textit{HTTPS link};
    \item[-] 1 Web-site has changed its password recovery mechanism from \textit{New Pw} to \textit{HTTPS Link};
    \item[-] 2 Web-sites do not work anymore, they were employing the \textit{HTTP link} mechanism and the \textit{HTTPS link} mechanism, respectively;
    \item[-] 1 Web-site (that was managing its password recovery with the \textit{New Pw} mechanism) no longer allows the user registration; and
    \item[-] 1 Web-site that was not working during the first analysis now work, and allows the registration of user accounts. This Web-site adopts the \textit{New Pw} password recovery mechanism.
\end{itemize}

\subsubsection{Spain} Even the Spanish Web-sites, as depicted in~\autoref{tab:old-analysis-vs-new-analysis}, have not made efforts to be compliant with the regulation. Indeed, the percentage of vulnerable Web-sites is exactly the same (i.e., 47.06\%). In detail, by the results of the new analysis, we note that:
\begin{itemize}
    \item[-] 15 Web-sites have not changed their password recovery mechanism;
    \item[-] 1 Web-site has changed its password recovery mechanism from \textit{New Pw} to \textit{HTTPS link};
    \item[-] 1 Web-site does not work anymore, it was employing the \textit{HTTPS link} mechanism; and
    \item[-] 1 Web-site that was not working during the first analysis now work, and allows the registration of user accounts. This Web-site adopts the \textit{HTTP link} password recovery mechanism.
\end{itemize}

\noindent The posthumous analysis to the entry into force of the GDPR shows how the Web-sites, although being among the more popular ones as they were occupying the top positions of Alexa's ranking in December 2017, continue to not comply with the regulation and risk incurring the heavy fines envisaged.

\subsection{Third Analysis: Top 200 Alexa's Web-sites of five European countries (December 2018)}
In this section, we provide an analysis of the top 200 Alexa's Web-sites of the five countries carried out in December 2018, a few months later the entry into force of the GDPR. The goal is to understand if the entry into force of the GDPR has improved the password management security of the Web-sites that were currently between the most popular according to the Alexa's ranking. As for the first analysis described in~\autoref{sec:first-analysis}, we first provide details about the filtering phase, in which we remove Web-sites demanding information we cannot produce, then we introduce the password recovery mechanisms adopted by the Web-sites considered.
\subsubsection{Filtering Phase} As described in subsection~\autoref{sec:methodology}, the Web-sites have been filtered to exclude both those that do not require authentication and the ones that demand information we are not able to produce. Of the 1,000 Web-sites, 687 do not require authentication---and hence have been excluded by our analysis---, while out of the remaining 313, we focus on 195---since 95 demanded information we cannot produce and 23 were not working during the analysis.

\begin{table}[h]
	\small
	\caption{Number of analyzed Web-sites among the top 200 per country (December 2018)}
	\label{tab:new-analyzed-websites}
	\centering
	\begin{tabular}{|C{3cm}|C{3cm}|}
		\hline
		\textbf{Country} & \textbf{Web-sites (\#)} \\
		\hline
		France & 34\\
		\hline
		Germany & 18\\
		\hline
		Italy & 38\\
		\hline
		Spain & 23\\
		\hline
		U.K. & 82\\
		\hline
		\textbf{Total} & \textbf{195}\\
		\hline
	\end{tabular}
\end{table}

The results of this filtering are shown in~\autoref{tab:new-analyzed-websites}: most of the Web-sites taken into account are from the U.K. (i.e., 82, approximately 42.05\%), while only a few Web-sites from Germany and Spain appear to require authentication from the user, 18 ($\approx$ 9.23\%) and 23 ($\approx$ 11.79\%), respectively. \\
We found interesting the result of the filtering process. Indeed, even if the intersection of the top 200 Alexa's Web-sites of the different countries in December 2017 and December 2018 is not null (i.e., from 35\% to 45\% of the considered Web-sites are different), it is possible to note that the percentage of the Web-sites that allows the authentication of users is almost the same. This could give some interesting ideas to characterize, at a regional level, the diffusion of Web-sites that allow users' registration.

\begin{table*}[h]
	\caption{Password recovery mechanisms adopted by the analyzed Web-sites (December 2018)}
	\label{tab:new-analysis}
	\centering
	\begin{adjustbox}{max width=\textwidth}
	\begin{tabular}{|C{1.5cm}C{1.5cm}|C{1.9cm}C{1.9cm}C{1.7cm}C{1.9cm}|C{1.75cm}}
	    \hline
		\textbf{Country} & \textbf{Web-sites (\#)}  & \textbf{Old Pw} & \textbf{New Pw} & \textbf{Temp Pw} & \textbf{HTTP link} & \multicolumn{1}{C{2cm}|}{\textbf{\% of vulnerable Web-sites}}\\
		\hline
		France & 34 & 0 & 9 & 0 & 3 & \multicolumn{1}{c|}{35.29}\\
		\hline
		Germany & 18 & 0 & 4 & 0 & 3 & \multicolumn{1}{c|}{38.88}\\
		\hline
		Italy & 38 & 8 & 10 & 1 & 3 & \multicolumn{1}{c|}{57.89}\\
		\hline
		Spain & 23 & 2 & 9 & 0 & 2 & \multicolumn{1}{c|}{56.52}\\
		\hline
		U.K. & 82 & 2 & 8 & 2 & 20 & \multicolumn{1}{c|}{32.02}\\
		\hline
		\textbf{Total} & \textbf{195} & \textbf{12 (6.15\%)} & \textbf{40 (20.5\%)} & \textbf{3 (1.54\%)} & \multicolumn{1}{c|}{\textbf{31 (15.9\%)}} &\\
		\cline{1-6}
	\end{tabular}
	\end{adjustbox}
\end{table*}

\noindent The results of the analysis are shown in~\autoref{tab:new-analysis} where, as done for~\autoref{tab:analysis-result}, each column represents a different method of password recovery adopted by the analyzed Web-sites---these password recovery methods can be observed from left to right in decreasing order of severity. Of the 195 Web-sites, 12 Web-sites ($\approx$ 6.15\%) employ the \textit{Old Pw} password recovery method, thus storing the passwords in clear, 40 Web-sites ($\approx$ 20.51\%) make use of the \textit{New Pw} password recovery mechanism, 3 Web-sites ($\approx$ 1.54\%) use the \textit{Temp Pw} password recovery mechanism, while 31 Web-sites ($\approx$ 15.9\%) allow users to recover the password through an \textit{HTTP link}, for a total of 44.12\% of vulnerable web-sites. \\
The analysis highlights how, even the Web-sites that are currently among the most popular after the entry into force of the GDPR, as they occupy the top positions of Alexa's ranking in the period of December 2018, are not compliant with the regulation, being subject to GDPR's almost unsustainable fines.

\section{Our Solution in a Nutshell} \label{sec:maildust}
The study discussed in~\autoref{sec:attackers-capabilities} show how adversaries who are able to access the e-mails of the users (e.g., mail service provider-level attackers and client intruder attackers) are threatening and can easily jeopardize both the privacy and security of the registered users. In this section, we discuss the details of \sol, an open-source module that could be adopted by any Web-site to provide registered users with a secure password recovery mechanism, preventing mail service provider-level attacks. \sol\ consists of two components, a server component, and a client component, respectively.\\

\subsection{\sol\ Server Module}
\begin{figure}[ht]
\centering
\includegraphics[scale=0.3]{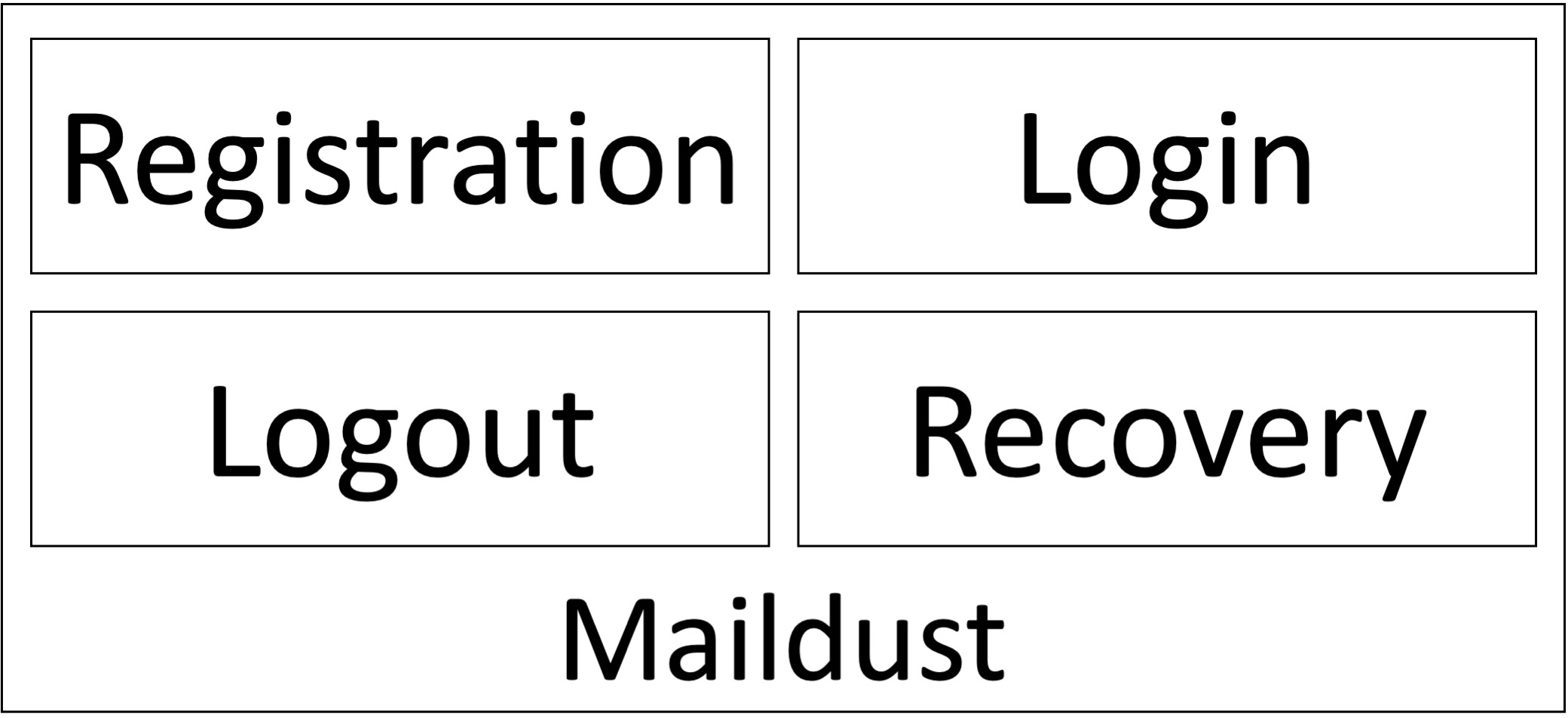}
\caption{\sol\ module}
\label{maildust-module}
\end{figure}

The \sol\ server module, depicted in~\autoref{maildust-module}, is composed of four submodules: Registration submodule, Login submodule, Logout submodule, and Recovery submodule, respectively.

\subsubsection{Registration} 
\begin{figure}[h]
\centering
\includegraphics[scale=0.8]{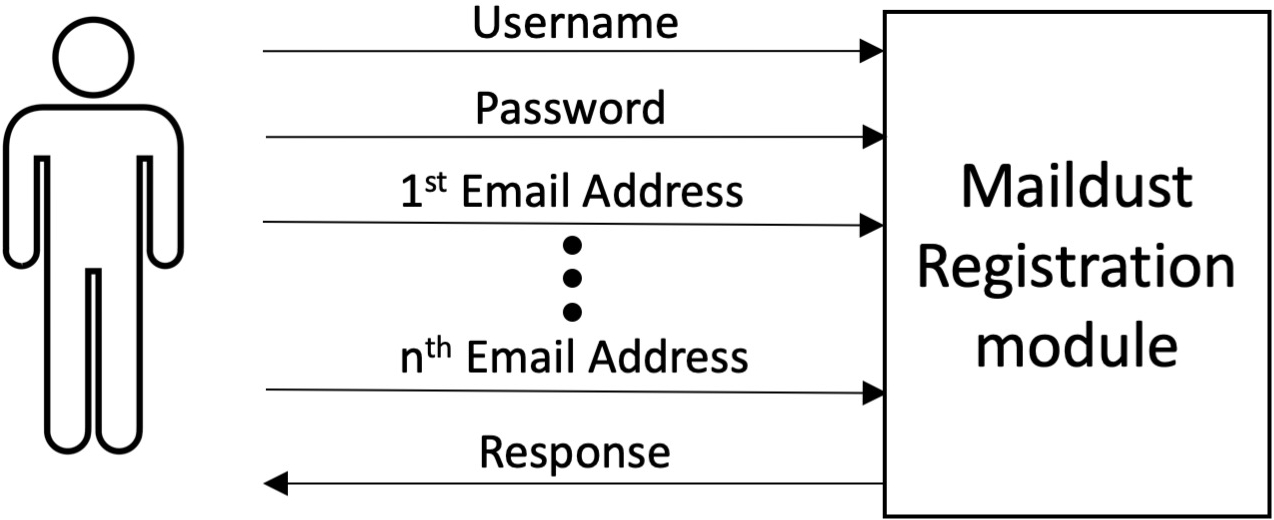}
\caption{Registration information}
\label{registration-1}
\end{figure}

The registration submodule allows users to authenticate themselves on any platform that integrates the \sol\ module. In this phase, depicted in~\autoref{registration-1}, the user will be required to provide authentication information, such as username, password, and several e-mail addresses. The e-mail addresses of the user will be used as password recovery resources.

\subsubsection{Login and Logout} The login and logout submodules allow users with a registered account to login to and logout from the platform that implements the \sol\ module.

\begin{figure*}[h]
\centering
\includegraphics[scale=0.23]{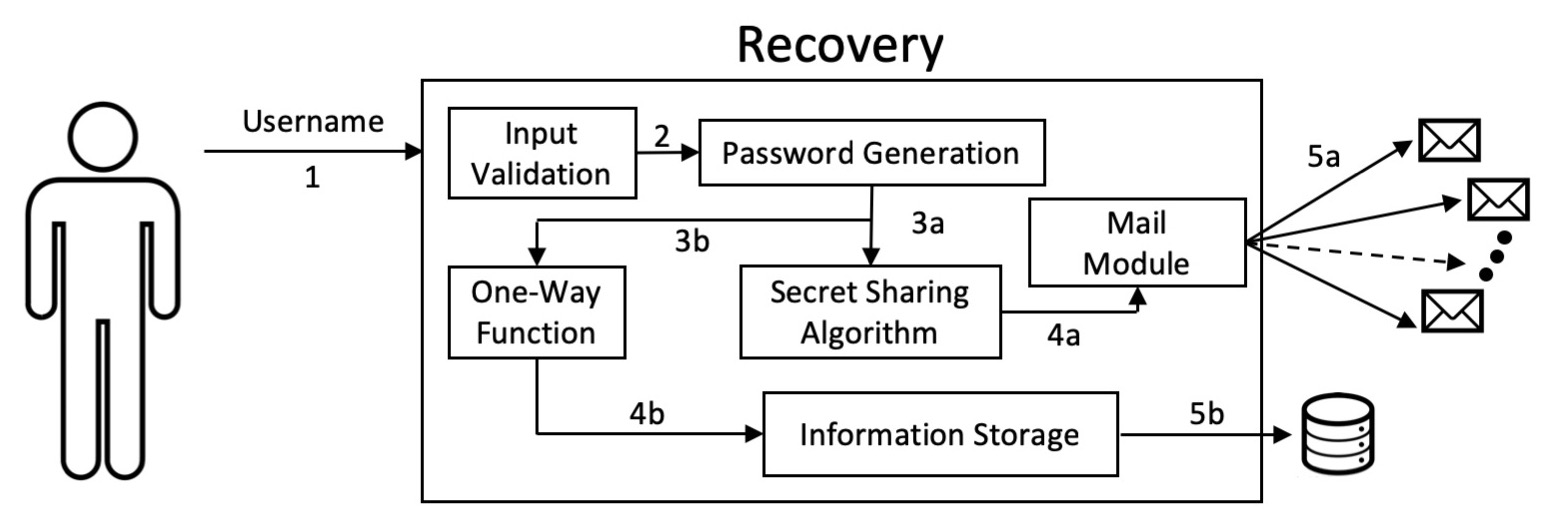}
\caption{Recovery submodule}
\label{recovery}
\end{figure*}

\subsubsection{Recovery} The recovery submodule allows users who have lost or forgotten their passwords to access the platform again. The recovery procedure, depicted in~\autoref{recovery}, consists of the following operations:
\begin{itemize}
    \item[1.] The user \textit{username}, that wants to recover her password, starts the recovery procedure from the \sol\ Recovery submodule.
    \item [2.] After verifying \textcolor{black}{that the user is currently} registered on the Web-site, the submodule generates a pseudorandom string, that will be assigned to the user (as a password).
    \item [3a.] A secret sharing algorithm (i.e., Shamir's secret sharing scheme) is applied to the password. The application of the Shamir's secret sharing scheme to the password allows to split it into N tokens such that a partial number K of tokens (K $\le$ N) will be enough to reconstruct the password. The total number of tokens N is given by the number of password recovery resources provided by the requesting user during the registration phase.
    \item [4a.] The tokens are sent to the mail module.
    \item [5a.] The mail module distributes the N tokens to the password recovery resources (i.e., e-mail addresses) provided by the users during the registration.
    \item [3b.] In parallel with the task 3a, a one-way function is applied to the password (i.e., SHA512).
    \item [4b.] The hashed password is sent to the information storage module.
    \item [5b.] The registration storage module associates the password with the requesting user, and stores the result inside a database.
\end{itemize}
The application of the one-way function allows \textcolor{black}{mitigating} possible attacks on the Web-site's database of registered users. Indeed, in case no operations are performed on the password before storing them, an attacker able to access the Web-site's database of registered users would read the users' password in clear. Although the passwords associated \textcolor{black}{with} the users are pseudorandom strings generated by the \sol\ submodule (i.e., associated with the user only on the Web-site in question), the attacker could easily obtain the credentials of all the users and impersonate them on the Web-site.

\subsection{\sol\ Client module}

\begin{figure*}[h]
\centering
\includegraphics[scale=0.90]{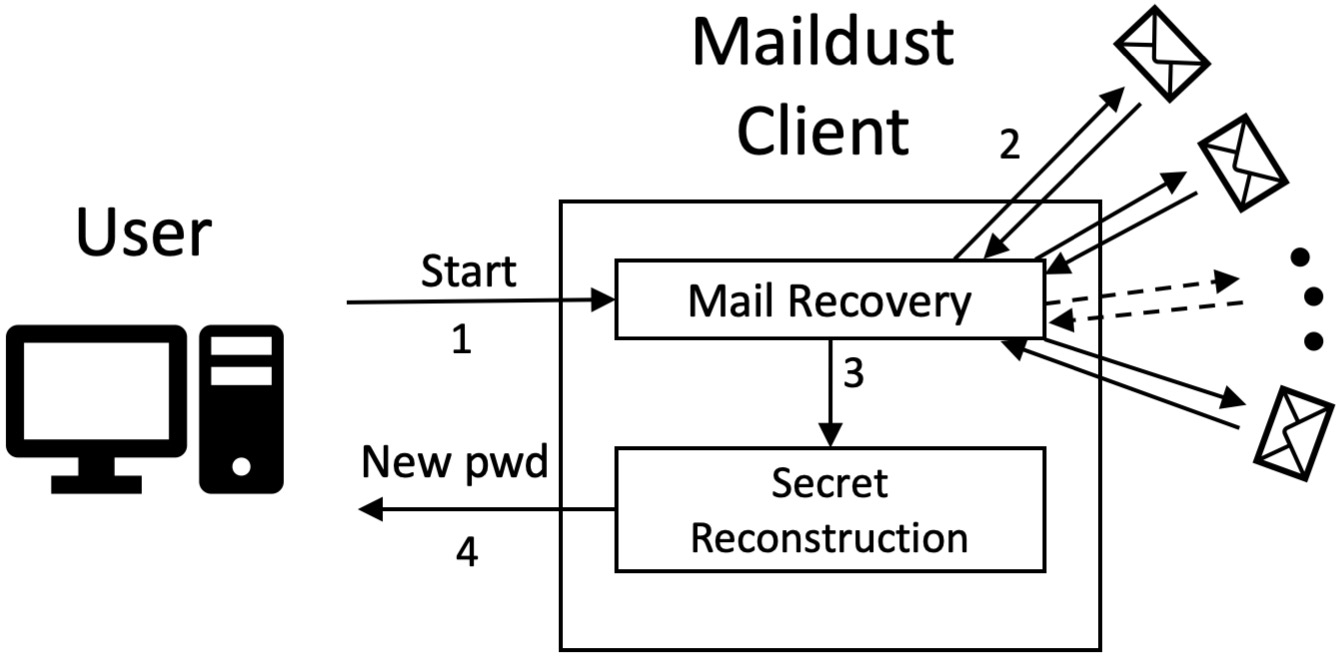}
\caption{Client module}
\label{client-module}
\end{figure*}

\noindent The \textcolor{black}{Maildust} client module, depicted in~\autoref{client-module}, is composed by two submodules: a mail recovery submodule and a secret reconstruction submodule, respectively.
\begin{itemize}
    \item[1.] The user, having started the \sol\ recovery procedure on a Web-site, requested the \sol\ client module to recover the new password.
    \item[2.] The Mail Recovery submodule allows \textcolor{black}{recovering} the tokens that have been distributed over the mail addresses of the user, provided as password recovery resources during the \sol\ registration phase.
    \item[3.] The tokens are sent to the Secret Reconstruction submodule, \textcolor{black}{which combine them to reconstruct the password} of the user. Similarly to the split of the password, the reconstruction of the password is 
    based on the Shamir's  scheme. Indeed, with just K tokens over N ($K \le N$) is it possible to  successfully reconstruct the password.
    \item[4.] The new password is returned to the user, that has the opportunity to access the Web-site again.
\end{itemize}



\subsection{Implementation} Given the modularity of the code and the possibility to easily integrate it in each Web-site, we decided to implement the \sol\ server module inside a Docker container. Containers technology, providing near bare-metal performance with respect to virtualization, is becoming a standard approach to distribute applications, allowing to run multiple versions of applications on the same machines~\cite{martin2018docker}. We decided to use Docker that, with more than 105 billion container downloads~\cite{about-docker}, is currently leading the market. The \sol\ Web server has been implemented over Flask~\cite{flask}, a lightweight microframework for Python, while for applying the Shamir's secret sharing scheme we relied on the open-source code published on GitHub, which is part of the secret-sharing project\footnote{https://github.com/blockstack/secret-sharing}. The client module has been implemented in Python 3.6 language, adopting libraries useful to manage the reading of received e-mails, such as imaplib\footnote{https://docs.python.org/3/library/imaplib.html\#module-imaplib} and email\footnote{https://docs.python.org/3/library/email.html}.

\begin{table}[t]
    \color{black}
    \caption{Password Recovery Mechanism Limitations}
    \label{tab:password_recovery_mechanism_requirements}
    \centering
    \begin{adjustbox}{max width=\textwidth}
    \begin{tabular}{|C{2.8cm}|C{2.8cm}|C{1.5cm}|C{1.2cm}|C{2.8cm}|}
    \hline
    \textbf{Pw Recovery Mechanism} & \textbf{Remember Security Details} & \textbf{Use more devices} & \textbf{Trust people} & \textbf{Trust the mail service provider} \\ \hline
    Security questions & \cmark & \xmark & \xmark & \xmark \\ \hline
    2-Factor Auth. & \xmark & \cmark & \xmark & \xmark \\ \hline
    Prev. used pw & \cmark & \xmark & \xmark & \xmark \\ \hline
    E-mail-based & \xmark & \xmark & \xmark & \cmark \\ \hline
    Vouching~\cite{brainard2006fourth}  & \xmark & \xmark & \cmark & \xmark \\ \hline
    \textbf{Maildust} & \xmark & \xmark & \xmark & \xmark \\ \hline
    \end{tabular}
    \end{adjustbox}
\end{table}

\textcolor{black}{\subsection{Usability} Our \sol\ solution,
besides being effective against mail-service provider level attacker, does not impact negatively on the usability and, therefore, on the user experience. Table~\ref{tab:password_recovery_mechanism_requirements} depicts the limitations 
for each of the password recovery mechanisms: 
\begin{itemize}
    \item \textbf{Security questions:} This mechanism requires the user to remember her security answers. There is a trade-off between the questions the user chooses and the level of security the mechanism can guarantee. Indeed, on the one hand the user makes no effort to remember trivial questions such as ``what was the name of your teacher in the primary school'' or ``what is your mother's maiden name''. However, on the other hand, the attacker could likely acquire such information in little time given the current level of social media exposure. This makes harder the choice of security questions, thus leading to the difficulty of remembering the answers.
    \item \textbf{2-Factor authentication: }The 2-Factor authentication mechanism requires two factors to authenticate a user. It takes into account both ``something the user knows'' and either ``something the user has (e.g., a physical token or a smart card)'' or ``something the user is'' (e.g., fingerprints or retinal scan). In the current implementations, Web-sites ask the user for a password (something she knows), and to confirm the access with her smartphone (something she has). Even if smartphones are increasingly widespread, there may be situations where 
    the user needs to recover her password and does not have the second device available.
    \item \textbf{Previously used passwords:} This mechanism requires the user to remember one (or a set of) previously used password(s). Besides its limitations (e.g., the user may not remember any of the previously used passwords), this password security mechanism presents possible severe security problems (e.g., informed attackers may know some of the previously used passwords of the victim).
    \item \textbf{E-mail-based recovery password:} This represents the most common password recovery mechanism currently adopted by the Web-sites. Once the password has been forgotten or lost, the user triggers the password recovery procedure and the new log-in information (e.g., the old password, a new password, a temp password, an HTTP/HTTPS link) are sent to the mail address of the user. This mechanism is effective provided there is trust in the mail-service provider. In our previous work~\cite{raponi2018spark}, we showed how the mail-service provider level attacker represents one of the most viable and dreadful attacks users are currently exposed to.
    \item \textbf{Vouching~\cite{brainard2006fourth}:} 
    This password recovery mechanism requires a registered user to rely on trustees, people previously appointed by the account holder, that will have the role of identifying her. These people will be contacted in case the password has been lost or forgotten, and they will be asked to perform the identification of the requesting user. This mechanism presents some availability issues (e.g., trustees may not be available in case of emergency) and privacy issues (e.g., there is the need to trust trustees).
    \item \textbf{\sol}: Our solution allows registered users to recover their passwords without remembering security details, neither requiring the users to rely on additional devices, nor trusting either people or mail service providers. In its simplest implementation, \sol\ just requires to install a program that allows automatically recovering the tokens the Web-site has sent to the different e-mails chosen by the account holder, followed by the automated password reconstruction, that is requiring a very mild cognitive load on the user side. \textcolor{black}{It is worth noting that the installation of the client module, although it improves the usability, is not strictly required to enjoy the benefits of \sol. Indeed, the \sol\ server module, instead of the password reconstructed by the \sol\ client module, could accept the tokens sent to the account holder by e-mail. Once received the tokens, the server module could apply the Shamir's secret sharing scheme algorithm to reconstruct the password, verify it, and allow the access to the user.} \\
    Finally, being the solution open-source, it is subject to public scrutiny and consequent improvement.
\end{itemize}
}

\section{Conclusion} \label{sec:conclusion}
In this paper, we have shed light on the weaknesses of  Web-based password management systems. The findings have been collected via a longitudinal study 
run  over some of the major EU Web-sites.
The resulting scenario sounds a loud alarm bell: some $28\%$ of the analyzed Web-sites are exposed to critical threats, while more than $44.12\%$ do show some severe vulnerabilities.
To cope with the above threat we have proposed \sol: a solution for password management and recovery that solves some of the critical issues current Web-sites are affected by. In particular, we have provided a thorough design and analysis of the proposed solution that, leveraging sound security primitives,
information spreading, and flexible tuning of the system parameters, allows users to tailor the solution to their perceived level of risk. \\*To the best of our knowledge, this is the first of its kind solution, and it comes with an open-source proof-of-concept.
We have shown how \sol\ do solve the above-cited security issues, and we believe that our decision to release the source code of our solution will allow researchers from both industry and academia to build on it, to enhance the degree of security and privacy of users of on-line Web-sites to the level they are entitled to.






\bibliographystyle{ACM-Reference-Format}
\bibliography{mybibfile}

\end{document}